\documentclass[twocolumn,preprintnumbers,amsmath,amssymbm,prd]{revtex4}
\usepackage{epsfig}
\usepackage{graphicx}

\begin{document}
\title{Higher-dimensional violations of the holographic entropy bound}
\author{Shahar Hod}
\affiliation{The Ruppin Academic Center, Emeq Hefer 40250, Israel}
\affiliation{ } \affiliation{The Hadassah Institute, Jerusalem
91010, Israel}
\date{\today}

\begin{abstract}
\ \ \ The holographic bound, $S\leq {\cal A}/4{\ell^2_P}$, asserts
that the entropy $S$ of a system is bounded from above by a quarter
of the area ${\cal A}$ of a circumscribing surface measured in
Planck areas. This bound is widely regarded as part of the elusive
fundamental theory of nature. In fact, the bound is known to be
valid for generic weakly gravitating isolated systems in {\it three}
spatial dimensions. Nevertheless, the entropy content of a physical
system is expected to be an increasing function of the number of
spatial dimensions (the more the dimensions, the more ways there are
to split up a given amount of energy). Thus, one may expect the
challenge to the holographic entropy bound to become more and more
serious as the number of spatial dimensions increases. In this paper
we explicitly show that thermal radiation in $D$ flat spatial
dimensions with $D\gtrsim 10^2$ may indeed violate the holographic
entropy bound.
\end{abstract}
\bigskip
\maketitle


\section{Introduction}

The influential holographic principle \cite{Hoof,Suss} asserts that
there is a deep relation between the physical content of a theory
defined in a spacetime and the corresponding content of another
theory defined on the boundary of the same spacetime \cite{Bek1}.
This principle is considered to be an important ingredient of the
ultimate physical theory of nature \cite{Bek1}.

The holographic principle suggests that the information encoded on
the boundary of a physical system should be able to describe the
entire set of possible quantum states of the bulk system
\cite{Bek1}. In light of the correspondence between information and
entropy \cite{inf1,inf2,inf3}, and the well-known entropy-area
relation for black holes \cite{inf3,Haw1}, this requirement has been
expressed in the form of the holographic entropy bound
\cite{Hoof,Suss,Bek1}:
\begin{equation}\label{Eq1}
S\leq {{\cal A}\over{4\ell^2_P}}\  ,
\end{equation}
where $\ell^2_P={G\hbar/c^3}$ is the Planck area. This bound thus
asserts that the entropy $S$ (or information) that can be contained
in a physical system is bounded in terms of the area ${\cal A}$ of a
surface enclosing it. It also implies that $(3+1)$-dimensional black
holes have the largest possible entropy among (stationary and
bounded) physical systems characterized by a given surface area
${\cal A}$ (see also \cite{Bek4,Hodspin,BekMayHod,Bek2}).

It should be mentioned that there do exist violations of the
holographic bound. For example, a collapsed object already inside
its own gravitational radius eventually violates it. The enclosing
area can only decrease while the enclosed entropy can only grow
\cite{Bous1,Bek1}. Another example is given by a large spherical
section of a flat Friedmann universe: its enclosing area grows like
radius squared while the enclosed entropy does so like radius cubed.
These examples belong to a class of strongly self-gravitating and
dynamical systems. The second example also describes an unisolated
system. A covariant entropy bound free from such failures was
introduced in \cite{Bous1}. However, this covariant bound is
characterized by a complicated formulation in cases which lack high
degree of symmetry \cite{Tav}. Thus, the original holographic bound,
which is simpler to apply, would still be very valuable if its range
of validity could be defined in a systematic way \cite{Bek1}.

Using the generalized second law (GSL) of thermodynamics, it was
shown \cite{Bek1} that the holographic bound (\ref{Eq1}) can indeed
be trusted for generic {\it weakly} self-gravitating {\it isolated}
systems in {\it three} spatial dimensions. In this paper we address
the following question: is the holographic bound also valid for
weakly self-gravitating isolated systems in {\it higher}-dimensional
spacetimes?

One may expect the challenge to the holographic entropy bound to
become more and more serious as the number of spatial dimensions
increases: the more the dimensions, the more ways there are to split
up a given amount of energy between the quantum states of the system
\cite{Bek3}. Thus, the entropy content of a physical system which is
characterized by a given amount of energy is expected to increase
with increasing number of spatial dimensions.

\section{$(D+1)$-dimensional radiation entropy}

Consider in $D$ flat spatial dimensions a spherical box of radius
$R$ into which we dump energy $E$ of massless fields. [We shall
henceforth use natural units in which $G=c=k_B=1$.] We shall follow
the analysis of \cite{Bek3} in order to calculate the system's
entropy in the thermodynamic regime. For the thermodynamic
description to be valid, the discreteness of the energy spectrum
(due to the finite size of the confining box) should be
unnoticeable. We shall therefore require that the characteristic
frequency of the thermal radiation be large compared to the energy
gaps which characterize the discrete energy spectrum of the bounded
system. This would imply that many wavelenghts small compared to $R$
are thermally excited. Below [see Eqs. (\ref{Eq14})-(\ref{Eq16})] we
shall make this statement more accurate.

The volume of a sphere of radius $R$ in $D$ spatial dimensions is
\begin{equation}\label{Eq2}
V_D(R)={{2\pi^{D/2}}\over{D\Gamma(D/2)}}R^D\  ,
\end{equation}
where $\Gamma(z)$ is the Euler gamma function. Consequently, the
volume in frequency space of the shell $(\omega,\omega+d\omega)$ is
\begin{equation}\label{Eq3}
dV_D(\omega)=D[V_D(\omega)/\omega]d\omega\  .
\end{equation}

The mean thermal energy in the sphere from one helicity degree of
freedom is given by
\begin{equation}\label{Eq4}
E=V_D(R){\int_0^{\infty}}{{\hbar\omega\
dV_D(\omega)}\over{{(e^{\beta\hbar\omega}\mp 1)}{(2\pi)^D}}}\  ,
\end{equation}
where the upper (lower) signs correspond to boson (fermion) fields,
and $\beta\equiv 1/T$ is the inverse temperature of the system. We
note that the distribution $\omega^{D}/(e^{\beta\hbar\omega}\mp 1)$
in Eq. (\ref{Eq4}) peaks at the characteristic frequency
\begin{equation}\label{Eq5}
\bar\omega={{D}\over{\hbar\beta}}[1\mp e^{-D}+O(e^{-2D})]\  .
\end{equation}

From Eqs. (\ref{Eq2})-(\ref{Eq4}) and the relation
\begin{equation}\label{Eq6}
{\int_0^{\infty}}{{x^Ddx}\over{e^x\mp1}}=\zeta(D+1)\Gamma(D+1)\times
\begin{cases}
1 & \text{for bosons} \ ; \\
1-2^{-D} & \text{for fermions} \ ,
\end{cases}
\end{equation}
where $\zeta(z)$ is the Riemann zeta function, one finds that the
mean energy of all massless fields is given by
\begin{equation}\label{Eq7}
E={{2N\zeta(D+1)\Gamma({{D+1}\over{2}})R^D}\over{{\pi}^{1/2}\Gamma({D\over
2})\beta^{D+1}\hbar^D}}\ ,
\end{equation}
where $N$ is the number of massless degrees of freedom (the number
of polarization states). Massless scalars contribute $1$ to $N$,
massless fermions contribute $1-2^{-D}$ to $N$ \cite{Bek3}, an
electromagnetic field contributes $D-1$ to $N$ \cite{CarCavHal}, and
the graviton contributes $(D+1)(D-2)/2$ to $N$ \cite{CarCavHal}.
Solving Eq. (\ref{Eq7}) for $\beta\hbar/R$ one finds
\begin{equation}\label{Eq8}
\beta\hbar/R=C_D(N\hbar/RE)^{{1}\over{D+1}}\  ,
\end{equation}
where
\begin{equation}\label{Eq9}
C_D\equiv\Big[{{2\zeta(D+1)\Gamma({{D+1}\over{2}})}\over{\pi^{1/2}\Gamma({D\over
2})}}\Big]^{{1}\over{D+1}}\ .
\end{equation}

Likewise, one can write the thermal entropy of one helicity degree
of freedom as \cite{Bek3}
\begin{equation}\label{Eq10}
S=V_D(R){\int_0^{\infty}} \Big[\mp \ln (1\mp
e^{-\beta\hbar\omega})+{{\beta\hbar\omega}\over{e^{\beta\hbar\omega}\mp
1}}\Big]{{dV_D(\omega)}\over{{(2\pi)}^D}}\ .
\end{equation}
After some algebra we obtain
\begin{equation}\label{Eq11}
S={{2N(D+1)\zeta(D+1)\Gamma({{D+1}\over{2}})R^D}\over{{\pi}^{1/2}D\Gamma({D\over
2})\beta^{D}\hbar^D}}\ ,
\end{equation}
which implies
\begin{equation}\label{Eq12}
S={{D+1}\over{D}}\beta E\  .
\end{equation}
Substituting Eq. (\ref{Eq8}) into Eq. (\ref{Eq12}) one finds
\begin{equation}\label{Eq13}
S=C_D(1+1/D)N^{1\over{D+1}}{(RE/\hbar)}^{D\over{D+1}}\
\end{equation}
for the $(D+1)$-dimensional radiation entropy.

\section{The thermodynamic criteria}

The description of our fixed energy system in terms of temperature
(the thermodynamic regime) is valid provided the characteristic
thermal frequency $\bar\omega$ [see Eq. (\ref{Eq5})] is large
compared to the characteristic energy gaps in the discrete spectrum
of the (finite size) system. [Otherwise, one cannot rely on
continuum formulae like Eqs. (\ref{Eq4}) and (\ref{Eq10}).] In the
familiar case of three spatial dimensions ($D=3$), this
thermodynamic condition implies that the dimensionless ratio
$\beta\hbar/R$ should satisfy the simple constraint $\beta\hbar/R\ll
1$. However, as we shall now discuss, the criterions for the
validity of the thermodynamic description in $D\gg 1$ spatial
dimensions are more involved: they depend on intricate combinations
of {\it two} dimensionless quantities: $\beta\hbar/R$ and $D$.

For a system in $D$ spatial dimensions confined into a region of
length-scale $R$, there are two distinct energy gaps which
characterize its discrete energy spectrum \cite{Notedisc}:
\begin{itemize}
\item{The characteristic energy gap between adjacent energy levels
is of the order of $\hbar/R$ \cite{Notedisc}. The validity of the
thermodynamic description requires this energy gap to be small
compared to the characteristic thermal frequency. Taking cognizance
of Eq. (\ref{Eq5}) for $\bar\omega$, one deduces the thermodynamic
condition
\begin{equation}\label{Eq14}
D/\beta\gg \hbar/R\  .
\end{equation}}
\item{The ground-state has a characteristic energy of the
order of $\sqrt{D}\hbar/R$ \cite{Notedisc}. The validity of the
thermodynamic description requires this ground-state energy to be
small compared to the characteristic thermal frequency. Thus, for
the thermodynamic description to be valid one should also have
\begin{equation}\label{Eq15}
D/\beta\gg \sqrt{D}\hbar/R\  .
\end{equation}}
\end{itemize}

The validity of the thermodynamic description also rests on the
assumption that {\it many} quanta (of each degree of freedom) are
thermally excited in the system: $E/N\gg \hbar\bar\omega$. Taking
cognizance of Eqs. (\ref{Eq5}) and (\ref{Eq7}), one finds the
condition
\begin{equation}\label{Eq16}
{C_D}^{D+1}{({{R}/{\beta\hbar}})}^D\gg D\  .
\end{equation}

For familiar physical systems in three spatial dimensions the three
requirements (\ref{Eq14})-(\ref{Eq16}) are basically identical.
(This may explain why, in the simple case of three spatial
dimensions, the various thermodynamic criteria are usually unified
into one requirement, $\beta\hbar/R\ll 1$.) However, for an
exceedingly large number of spatial dimensions ($D\gg 1$) the three
requirements are not identical -- in fact, condition (\ref{Eq16})
enforces the strongest constraint. Taking cognizance of Eq.
(\ref{Eq8}), one finds that the strongest thermodynamic condition
may be cast in the form
\begin{equation}\label{Eq17}
{C^{-1}_D}{(N\hbar/RE)}^{D\over{D+1}}\ll D^{-1}\  .
\end{equation}
We characterize this constraint by the dimensionless control
parameter $\xi$ defined by
\begin{equation}\label{Eq18}
\xi\equiv C^{-1}_D D{(N\hbar/RE)}^{D\over{D+1}}\ll 1\  .
\end{equation}
Below we shall come back to this thermodynamic condition.

\section{The weak-gravity criteria}

Our analysis is appropriate only for weakly self-gravitating
systems. In particular, formula (\ref{Eq13}) for the system's
entropy can be trusted only in a limited range of parameters
(limited range of energies for a given system's radius $R$) as here
stated. The spacetime outside the spherical box (for $D\ge 3$) is
described by the $(D+1)$-dimensional Schwarzschild-Tangherlini
metric \cite{SchTang,Kun} of ADM energy $E$:
\begin{equation}\label{Eq19}
ds^2=-H(r)dt^2+{H(r)}^{-1}dr^2+r^2d\Omega^{(D-1)}\ ,
\end{equation}
with
\begin{equation}\label{Eq20}
H(r)=1-{\Big({r_g\over r}\Big)}^{D-2}\  .
\end{equation}
Here
\begin{equation}\label{Eq21}
r_g={\Big[{{16\pi E}\over{(D-1)A_{D-1}}}\Big]}^{1\over{D-2}}
\end{equation}
is the gravitational radius of the box and
\begin{equation}\label{Eq22}
A_{D-1}={{2\pi^{D/2}}\over {\Gamma(D/2)}}
\end{equation}
is the area of a unit $(D-1)$-sphere.

For the system to be weakly self-gravitating, one should impose the
criterion $H(r=R)\simeq 1$ at the surface of the sphere, or
equivalently $(r_g/R)^{D-2}\ll 1$. Taking cognizance of Eqs.
(\ref{Eq20})-(\ref{Eq21}), this condition yields the restriction
\begin{equation}\label{Eq23}
RE\ll {{D-1}\over{16\pi}}{\cal A}\  ,
\end{equation}
where ${\cal A}=A_{D-1}R^{D-1}$ is the surface area of the system.
We characterize this restriction by the dimensionless control
parameter $\eta$ defined by
\begin{equation}\label{Eq24}
\eta\equiv {{16\pi RE}\over{(D-1){\cal A}}}\ll 1\  .
\end{equation}

A similar (but somewhat weaker) constraint can be obtained from the
requirement that the magnitude of the system's interior
energy-momentum tensor should be much less than the scalar curvature
at the surface. For a physical system confined by a
$(D+1)$-dimensional sphere this requirement reads:
\begin{equation}\label{Eq25}
{E\over{V_D(R)}}\ll {{(D-1)(D-2)}\over {R^2}}\  .
\end{equation}
Using the relation $V_D(R)=A_{D-1}R^{D-1}\times {R\over D}$, one
obtains the constraint
\begin{equation}\label{Eq26}
RE\ll {{(D-1)(D-2)}\over{D}}{\cal A}\  .
\end{equation}
This criterion is in the same spirit of (\ref{Eq23}).

Taking cognizance of Eqs. (\ref{Eq13}) and ({\ref{Eq24}), we can
write the system's thermal entropy as
\begin{equation}\label{Eq27}
S=C_D(1+1/D)N^{1\over{D+1}}{\Big[{{\eta(D-1){\cal
A}}\over{16\pi\hbar}}\Big]}^{D\over{D+1}}\ .
\end{equation}
We shall now address the following question: can the
$(D+1)$-dimensional radiation entropy exceed the holographic entropy
bound (\ref{Eq1})?

\section{The holographic bound}

For the system's entropy (\ref{Eq27}) to beat the holographic bound
(that is, $S>A/4\hbar$), its surface area must be bounded from above
according to
\begin{equation}\label{Eq28}
{{\cal
A}\over\hbar}<{[4C_D(1+1/D)]}^{D+1}N{\Big[{{\eta(D-1)}\over{16\pi}}\Big]}^{D}\
.
\end{equation}
Solving Eqs. (\ref{Eq18}) and (\ref{Eq24}) for $RE$, one can express
the system's area as
\begin{equation}\label{Eq29}
{{\cal A}\over{\hbar}}={{16\pi N D^{{D+1}\over
D}}\over{\eta(D-1)(\xi C_D)^{{D+1}\over D}}}\  .
\end{equation}
Substituting (\ref{Eq29}) into (\ref{Eq28}), we realize that a
violation of the holographic bound can only occur if the number of
spatial dimensions satisfies the inequality \cite{Notecd}
\begin{equation}\label{Eq30}
D\ge D^*\simeq 4\pi/\eta\xi^{1/D}\  .
\end{equation}
Since $\xi\ll 1$ and $\eta\ll 1$ [see Eqs. (\ref{Eq18}) and
(\ref{Eq24})], we learn from (\ref{Eq30}) that thermal radiation in
three spatial dimensions conforms to the holographic bound.

In order to estimate the critical dimension $D^*$ (the minimal value
of $D$ above which a violation of the holographic bound can be
realized) one may substitute $\eta_{\text{max}}=O({10}^{-1})$ and
$\xi_{\text{max}}=O({10}^{-1})$ for the dimensionless control
parameters. We then find that the entropy of our $(D+1)$-dimensional
system may exceed the holographic bound for
\begin{equation}\label{Eq31}
D\ge D^*=O(10^2)\  .
\end{equation}

It should be recognized that the precise value of the critical
dimension $D^*$, Eq. (\ref{Eq31}), can be challenged. This is a
direct consequence of the inherent fuzziness in the numerical values
of the control parameters $\eta_{\text{max}}$ and
$\xi_{\text{max}}$. This is the price we must pay for not giving our
problem a full quantum treatment in curved spacetimes. Nevertheless,
it should be clear that there is some critical value for the number
of spatial dimensions [see Eq. (\ref{Eq30})], probably around
$D^*\sim 10^2$, above which the system's entropy exceeds the
holographic bound.

Are there any relevant effects which might change our conclusion? We
note that thermal fields restricted to a finite region exert
pressure on the region's boundary. This pressure must be balanced by
a tensile wall or "container", whose mass $E_{\text{box}}$ should be
added to the energy $E$ of the confined fields \cite{BekBous}. The
container must be sufficiently rigid to withstand the pressure
caused by the thermal fields. A lower bound on the surface density
$\sigma$ (energy per surface area) of the required container was
obtained in \cite{BekBous} for the case of three spatial dimensions:
\begin{equation}\label{Eq32}
\sigma\geq {{p}\over{\hat K}}\  ,
\end{equation}
where $p$ and $\hat K$ are the pressure exerted by the enclosed bulk
system on the container and the trace of the extrinsic curvature,
respectively. (The only non-vanishing entries of $K_{ab}$ are the
spatial components tangential to the container \cite{BekBous}.) This
result can readily be generalized to higher-dimensional spacetimes.
Thus, for a spherical box in $D$ spatial dimensions one finds
\begin{equation}\label{Eq33}
E_{\text{box}}\geq {{{\cal A}p}\over{\hat K}}\
\end{equation}
for the energy of the confining box. Substituting $p={1\over
D}[E/V_D(R)]$ and $\hat K=(D-1)/R$ in (\ref{Eq33}), and using the
relation $V_D(R)={\cal A}\times {R\over D}$, one finds that the
minimal container's mass, $E_{\text{box}}^{\text{min}}$, satisfies
the relation
\begin{equation}\label{Eq34}
{{E_{\text{box}}^{\text{min}}}\over{E}}={1\over{(D-1)}}\  .
\end{equation}
Thus, the contribution of the container's mass to the total energy
of the system can be made negligible in the large $D$ limit
(\ref{Eq31}).

\section{The generalized second law of thermodynamics}

It was shown \cite{Bek1} that the holographic bound for generic
weakly self-gravitating isolated systems in three spatial dimensions
is a direct outcome of the generalized second law (GSL) of
thermodynamics. So, does the above violation of the holographic
entropy bound (by weakly self-gravitating physical systems in {\it
higher}-dimensional spacetimes) imply a violation of the GSL?

Consider a gedanken experiment in which a $(D+1)$-dimensional
spherical object with negligible self-gravity is deposited at the
horizon of a $(D+1)$-dimensional black hole with the least possible
energy. A violation of the GSL seems to occur unless the system's
entropy $S$ is bounded from above according to the universal entropy
bound \cite{Bek4,Bous1}:
\begin{equation}\label{Eq35}
S\leq 2\pi RE/\hbar\  .
\end{equation}
Arguing from the GSL, \cite{Bek4} has proposed the existence of the
entropy bound (\ref{Eq35}). The entropy of the swallowed object
disappears, but an increase in black-hole entropy occurs which
guarantees that the GSL is respected (the total entropy of black
hole + object never decreases) provided the object's entropy $S$ is
bounded by (\ref{Eq35}).

Now, taking cognizance of the thermodynamic condition (\ref{Eq18}),
one realizes that the system's thermal entropy (\ref{Eq13}) may be
expressed as
\begin{equation}\label{Eq36}
S=C_D(1+1/D)(\xi C_D/D)^{1/D}{RE/\hbar}\ .
\end{equation}
From Eq. (\ref{Eq36}) with $\xi\ll 1$ one deduces $S<RE/\hbar$ for
all finite values of $D$. In fact, $S\to RE/\hbar$ in the
$D\to\infty$ limit. The system's thermal entropy therefore conforms
to the entropy bound (\ref{Eq35}) regardless of the value of  $D$.
Thus, the GSL is {\it respected} despite the fact that the entropy
of the system may exceed the holographic bound (\ref{Eq1}).

\section{Summary}

It is well-established that the holographic bound (\ref{Eq1}) can be
trusted for generic weakly self-gravitating isolated systems in {\it
three} spatial dimensions. However, in this paper we have shown that
this is {\it not} a generic property of the holographic bound:
weakly self-gravitating isolated thermal radiation in $D\gtrsim
10^2$ spatial dimensions may violate the holographic entropy bound.
There is obviously no evidence that the actual number of spatial
dimensions in nature is so large. In this sense, the breakdown of
the holographic bound is not manifest empirically. In fact, arguing
from the holographic principle, one may {\it conjecture} that the
seeming clash between the holographic bound and an exceedingly large
number of spatial dimensions merely tells us that physics is
consistent only in a world with a limited number of large spatial
dimensions, such as ours.

\bigskip
\noindent
{\bf ACKNOWLEDGMENTS}

This research is supported by the Meltzer Science Foundation. I
thank Liran Shimshi, Clovis Hopman, Celina Cohen-Saidon, Yael Oren,
and Arbel M. Ongo for helpful discussions. I thank Jacob D.
Bekenstein for helpful correspondence.



\begin{thebibliography}{99}
\bibitem{Notedisc} This is most easily seen in the example of a massless
scalar field confined to a $(D+1)$-dimensional cubic cavity of side
length $L$, with Dirichlet boundary conditions. The energy spectrum
is given by ${\cal E}={{\hbar\pi}\over
L}(\sum_{i=1}^{D}{k_i}^2)^{1/2}$, where the $\{k_i\}_{i=1}^D$ run
over the positive integers. Thus, the ground-state energy is ${\cal
E}_{\text{ground}}={{\hbar\pi\sqrt{D}}\over{L}}$ and the
characteristic energy gap between adjacent energy levels is
$\Delta{\cal E}=O({{\hbar\pi}\over{L}})$.
\end{thebibliography}
\end{document}